\documentclass[11pt]{article}

\usepackage[a4paper, margin=1in]{geometry}

\usepackage{mathptmx}
\usepackage[T1]{fontenc}
\usepackage[utf8]{inputenc}

\usepackage{amsmath, amssymb, amsthm}
\usepackage{graphicx}
\usepackage{booktabs}
\usepackage{microtype}
\usepackage[hidelinks]{hyperref}
\usepackage{url}

\usepackage{titlesec}
\titleformat{\section}{\large\bfseries}{\thesection}{0.6em}{}
\titleformat{\subsection}{\normalsize\bfseries}{\thesubsection}{0.6em}{}
\titlespacing*{\section}{0pt}{1.4ex plus .2ex}{0.8ex}

\usepackage[numbers,sort&compress]{natbib}
\title{\vspace{-1.2cm}\bfseries\LARGE Exploring the Cryptographic Limits of Transformer Networks}
\date{}  

\author{%
  \large
  Stefan Domunco\textsuperscript{1}\quad
  Andis Draguns\textsuperscript{2}\quad
  Philip Torr\textsuperscript{1}\\[1.0ex]
  Isaac Robinson\textsuperscript{1}\quad
  Christian Schroeder de Witt\textsuperscript{1}\\[1.2ex]
  \normalsize
  \textsuperscript{1}University of Oxford\qquad
  \textsuperscript{2}Contramont Research\\[0.6ex]
  \texttt{stefan.domunco@cs.ox.ac.uk}
}

\usepackage{algorithm}
\usepackage{algpseudocode}

\begin{document}
\maketitle

\renewcommand{\thefootnote}{}\footnotetext{Preprint. Under review.}\renewcommand{\thefootnote}{\arabic{footnote}}

\begin{abstract}
\noindent
In recent work it has been shown that colluding AI agents can use steganographic 
methods to exchange malicious information~\cite{motwani2025secretcollusion}. Whether a 
transformer can implement steganographic methods depends on what cryptographic 
functions it can implement, since a transformer that can implement a cryptographic 
function within its layers has source-free randomness access (Theorem~3,~\cite{motwani2025secretcollusion}). 
Despite existing circuit-complexity results, no prior work maps specific 
cryptographic constructions to transformer architectures. As Merrill et al.~\cite{merrill2022saturated} 
have shown that saturated transformers can be seen as threshold circuits, we first 
generate threshold circuits for three different cryptographic constructions (Keccak 
functions, Merkle--Damg\aa rd constructions and Merkle Trees) and then map these 
circuits to different transformer architectures. We derive verified scaling laws 
for the width and depth of the circuits which implement each cryptographic 
construction and propose two different mappings: no-attention mapping, tokens-as-gates mapping.
Beyond its security implications, this work contributes to \emph{eval science}
by establishing a methodology for deriving structural guarantees on
transformer computational capacity. Specifically, we derive constructive upper bounds on
what a transformer of a given depth and width could plausibly compute, providing
a principled foundation for capability evaluations of transformer-based AI systems.
\end{abstract}

\vspace{1ex}


\section{Introduction}
\label{sec:introduction}
Autonomous systems such as LLMs have become widely deployed across applications in which AI agents interact directly, including multi-agent robotics~\cite{OpenXEmbodiment} and trading strategies~\cite{OilGasMARL}. 
Previous
work~\cite{motwani2025secretcollusion} shows that interacting AI agents can send each other sensitive
information in plain sight, using messages that look harmless to an observer via steganography ~\cite{cachin2004steganography}. Moreover, it has been shown that backdoors can be compiled into
transformers~\cite{draguns2024unelicitable} such that they cannot be detected by polynomial-time
algorithms using activation-based interpretability methods. Such backdoors could be triggered by steganographic inputs from a colluding agent, or an external adversary.
This raises
the concern that AI agents could covertly exchange sensitive information with one another, posing an unmonitorable security risk.
Crucially, in Theorem 3 of~\cite{motwani2025secretcollusion} this is shown to require only that the model can compute a cryptographic hash function, making transformer cryptographic expressivity a direct security question.

Merrill et al.~\cite{merrill2022saturated} show that saturated transformers can
be represented as circuits of class $\mathrm{TC}^0$, establishing circuit complexity as a natural
framework for reasoning about the computational limits of transformer architectures (see Section \ref{bg:sat_trans}). 

While prior work establishes both the circuit-theoretic limits of saturated transformers and the
security concerns that interacting AI agents pose, no concrete work maps cryptographic constructions
such as Keccak functions, MDCs and MTs (see Sections \ref{bg:keccak}, \ref{bg:MD}, \ref{bg:MT}) to transformer
architectures using their circuit representations. This paper provides the first systematic circuit-level analysis of these constructions in the context of transformer expressivity and proposes concrete mappings from these circuits to transformers.
Concretely, the contributions of this work are:
\begin{itemize}
\item We derive and verify scaling laws for the depth and width of threshold circuits implementing Keccak functions, Merkle--Damg\aa rd constructions, and Merkle Trees.
\item We propose two concrete mappings from these circuits to transformer architectures, establishing constructive upper bounds on the required depth and FFN width.
\item This work lays the foundations for a novel class of AI safety evaluations~\cite{liang2023helm} driven by the computational capacity of the agent, offering structural guarantees on what a transformer-based system could plausibly compute.
\end{itemize}

\section{Preliminaries}
\label{sec:background}

\subsection{Transformers}
Vaswani et al.~\cite{vaswani2017attention} define a transformer as a stack of $N$ identical
layers, each pairing a self-attention sublayer with a position-wise feed-forward
network (FFN). We use only the encoder stack. With queries, keys, and values
packed into $Q$, $K$, $V$ (queries and keys of dimension $d_k$), attention is
\begin{equation}
\mathrm{Attention}(Q,K,V) = \mathrm{Softmax}_{\mathrm{row}}\!\left(\frac{QK^{\top}}{\sqrt{d_k}}\right)V,
\end{equation}
where $\mathrm{Softmax}_{\mathrm{row}}$ normalises each row into a probability
distribution. We instantiate each FFN with
the SwiGLU gating of Shazeer~\cite{shazeer2020glu}, matching the MLPs compiled by the
Reifier compiler (Section~\ref{sec:methodology}).

\subsection{Circuits}
 
A boolean circuit is a directed acyclic graph of $n \in \mathbb{N}$ boolean inputs (sources) and $m \in \mathbb{N}$ boolean outputs (sinks). Each input is represented by a vertex with no incoming edges and each intermediate node is called a gate. Vertices with no incoming edges that are not sources are constants with a fixed value 0 or 1. Thus we can see a circuit as a function of boolean inputs and boolean outputs $f : \{0,1\}^n \to \{0,1\}^m$.
 
In 1993 Hajnal et al. ~\cite{hajnal1993threshold} defined the threshold gate for threshold \(\theta \in \mathbb{Z}\) and integer weights \(\alpha \in \mathbb{Z}^m\) for input \(x \in \{0, 1\}^m\) as follows:
\begin{equation}T(x) = \begin{cases} 1 & \text{for } \sum_{i=1}^m \alpha_i \cdot x_i \geq \theta \\0 & \text{otherwise} \end{cases}\end{equation}   
 
The depth of a circuit is defined as the longest path from an input to an output and the width of a circuit is equal to the maximum number of gates in a circuit layer.
 
Hajnal et al. (1993)~\cite{hajnal1993threshold} define the $\mathrm{TC}^i$ class as the class of decision problems decidable by uniform Boolean circuits with depth $O((\log n)^i)$ and polynomial size in the input using only threshold gates with unbounded fan-in. This generalizes to $\mathrm{TC} = \bigcup_i \mathrm{TC}^i$.
 
\subsection{Cryptographic Primitives}

\subsubsection{Collision Resistant Hash Function (CRHF)}

A collision-resistant hash function (CRHF) is pre-image, second-pre-image, and collision resistant: its outputs cannot be inverted, no second input matching a given output can be found, and no two distinct inputs produce the same output. A hash function $h$ is a compression function if it maps a fixed-size input to a smaller fixed-size output.

\subsection{Merkle--Damg{\aa}rd Construction (MDC)}
\label{bg:MD}

Independently in 1989, Merkle~\cite{merkle1989one} and Damgård~\cite{damgard1989design} presented the same construction.
Both constructions build a CRHF using a smaller
CRHF compression function. Merkle proposed a method to build a CRHF
$F:\{0,1\}^* \rightarrow \{0,1\}^o$ from a CRHF compression function
$F_0: \{0,1\}^{m} \rightarrow \{0,1\}^o$. $F$ can be computed from $F_0$
as follows:

\begin{algorithm}[h]
\caption{Merkle--Damg{\aa}rd construction}
\begin{algorithmic}[1]
\Function{F}{$x$ -- arbitrary sized input split into $n$ same sized chunks}
    \State $\mathit{result} \gets 0$
    \For{$i = 1$ \textbf{to} $n$}
        \State $\mathit{result} \gets F_0(\mathit{result},\, x[i])$
    \EndFor
    \State \Return $\mathit{result}$
\EndFunction
\end{algorithmic}
\end{algorithm}

Before this construction, Merkle  pads the
input to a multiple of $b_{\mathit{size}}$ which is the block size used by
$F_0$. Define $c_{\mathit{size}}$ as the size of the intermediate result and
call it chain size and then we have the equation
$b_{\mathit{size}} + c_{\mathit{size}} = m$. Since different inputs can pad
to the same $x$, a second padding block is appended, encoding the length of the original message in binary. For $x_0$ the
original input, $\mathit{pad}_0$ the padded 0s and $\mathit{pad}_{\mathit{len}}$
the additional padding for the length we get that
$x = x_0 \| \mathit{pad}_0 \| \mathit{pad}_{\mathit{len}}$
where $\|$ represents concatenation.
 
\subsection{Merkle Trees}
\label{bg:MT}
 
The Merkle Tree (MT) construction~\cite{merkle1989certified} follows a Divide and Conquer strategy that
hashes $n$ documents together in a tree structure.
We can assume that each document is a block of text from the message
that we want to hash.
First, ordering the message blocks, define $H(i,j)$ to be the hash
of the blocks in the interval $[i,j]$ using the recurrence:
 
\begin{align}
    H(i,\,i) &= F(\text{block}_{i}) \label{eq:merkle_base} \\[6pt]
    H(i,\,j) &= F\!\left(
                    H\!\left(i,\,\left\lfloor\tfrac{i+j}{2}\right\rfloor\right)
                    \;\|\;
                    H\!\left(\left\lfloor\tfrac{i+j}{2}\right\rfloor+1,\,j\right)
                  \right) \label{eq:merkle_rec}
\end{align}

Here $F$ is a compression CRHF.

\subsection{Keccak Functions}
\label{bg:keccak}
The sponge construction~\cite{sponge_duplex} is an iterative approach
which works in 2 similar phases: absorb and squeeze. The absorb phase combines the information
from the whole input sequentially, keeping a current state $a$ at all
times which represents the combined data. The state has 2 parts: the capacity $c$
and the rate $r$. The rate represents the bits that we can read and write
from the state and the capacity represents the part of the state that we cannot access. Before
the absorb phase, we pad the input such that it has length a multiple of $r$ and then split it
into equally sized blocks. Moreover, we define $f$ to be the block function which combines the
information of the current block with the information of the state at each step of the absorb and squeeze phases. We define the
absorb phase in Algorithm 2. The squeeze phase generates the output and we describe it in Algorithm 3. Both algorithms are presented in Appendix \ref{app:keccak_def}.

The Keccak function is a family of sponge functions in which $f$ is performed by 5 different
operations: $\theta, \rho, \pi, \chi, \iota$ and a specific padding function $pad$ is used. These functions and other Keccak definitions are presented in Appendix \ref{app:keccak_def}.

\subsection{Saturated Transformers}
\label{bg:sat_trans}

A \textit{saturated} transformer~\cite{merrill2022saturated} distributes attention uniformly across the elements
with maximum pre-softmax attention score:

\begin{align}
    \mathcal{M}(a) &= \{i \mid a_i = \max_j\, a_j\} \\[6pt]
    \mathrm{saturated\_attention}(a)_j &=
    \begin{cases}
        \dfrac{1}{|\mathcal{M}(a)|} & j \in \mathcal{M}(a) \\[6pt]
        0 & \text{otherwise}
    \end{cases}
\end{align}        

\section{Methodology}
\label{sec:methodology}

Following Merrill et al.~\cite{merrill2022saturated}, whose result is summarised in Section~\ref{sec:background}, we first map our cryptographic constructions to threshold circuits and then to saturated transformers.
The \texttt{Reifier} library~\cite{reifier} has been used to compile given boolean functions into layered threshold circuits, implemented as MLPs with SwiGLU-gated layers. For a function $f:\{0,1\}^n \to \{0,1\}^m$, instead of computing $f$, it traces each bit operation (\textsc{and}, \textsc{or}, \textsc{not}) into threshold gates, producing a DAG with $n$ input nodes and $m$ sink nodes that follows the computational flow. To satisfy the layered MLP structure, in which each node depends only on the immediately preceding layer, \texttt{Reifier} inserts identity nodes. The main limitation is that this produces wide MLPs when values must be carried across many non-consecutive layers.

The experiments have been run on Keccak functions, MDCs and MTs. These were chosen because they represent three different classes of hashing methods and are widely deployed, making their analysis practically relevant.          

\section{Scaling Laws}
\label{sec:experiments}

\subsection{Keccak Experiments and Observations}
\label{subsec:keccak}

The experiments were run on toy versions of Keccak with small $w$ and $n_{\text{rounds}}$ values
in order to understand the overall topology of the circuit.

\subsubsection{Layered Circuit Interpretation}
\label{subsubsec:keccak-circuit}

The first experiment was conducted on an input of 17 bits, $w=1$ and $n_{\text{rounds}}=1$. The detailed compiled circuit can be found in \texttt{keccak\_logw1\_round1.txt} in the
accompanying GitHub repository~\cite{github}. Figure \ref{fig:keccak1} in Appendix \ref{app:keccak_experiments} presents a simplified diagram of the circuit.

This experiment established how Keccak rounds map to circuit layers. Padding was not necessary for this experiment.

\subsubsection{Depth Scaling}
\label{subsubsec:keccak-depth}

From the compiled circuit we can establish the depth and width scaling
laws for a single-block Keccak circuit. One complete round consists of six layers
($\theta$: 2, $\chi$: 3, $\iota$: 1; $\rho$ and $\pi$ are free), plus one input
layer and one output layer, giving
\begin{equation}
    \mathrm{depth}_{\mathrm{circuit}} = 6n_{\text{rounds}} + 2,
\end{equation}
where the additive constant 2 accounts for the input and output layers. The widest
layer is the first $\theta$ threshold layer, giving
\begin{equation}
    \mathrm{width}_{\mathrm{circuit}} = 11 \cdot 25w.
\end{equation}
The linear prediction $\mathrm{depth}_{\mathrm{pred}}(n_{\text{rounds}}) = 6n_{\text{rounds}} + 2$
holds for most entries. Occasional deviations, shown in Table~\ref{tab:keccak-pred-vs-obs},
arise because the $\iota$ layer is omitted when the corresponding round constant is
zero; this is an artefact of the compiler rather than a property of the construction
itself.

\subsubsection{Extension to Arbitrary-Length Inputs}
\label{subsubsec:keccak-arbitrary-input}

The results above generalise to arbitrary-length inputs. The only difference is that
the unused block bits must be carried through the circuit. Let $p$ denote the size
of the padded input message and $n_b$ the number of blocks it is split into. The
maximum width is reached in the first $\theta$ threshold layer of the first round,
where all unabsorbed message bits must be stored alongside the full Keccak state:
\begin{equation}
    \mathrm{width}_{\mathrm{circuit}} = 11 \cdot 25w + p - r.
\end{equation}
For depth, XOR-ing the current block into the $r$-bit rate portion of the state
requires two additional layers per block, giving
\begin{equation}
    \mathrm{depth}_{\mathrm{circuit}} = 2 + n_b(2 + 6n_{\text{rounds}}).
\end{equation}

\subsubsection{Arbitrary Output Length}
\label{subsubsec:keccak-arbitrary-output}

To capture the full sponge behaviour, in which Keccak maps an arbitrary-length input
to an arbitrary-length output, a further experiment was conducted with $\log w = 2$,
$n_{\text{rounds}} = 1$, one input block, an output of 64 bits, and $r = 25$. During the
squeeze phase the current output bits must also be carried until the computation
completes, introducing a second bottleneck for the width. The width formula becomes
\begin{equation}
    \mathrm{width}_{\mathrm{circuit}}
    = \max\!\left(11 \cdot 25w + p - r,\;
                  25w + r\left\lceil\frac{o-r}{r}\right\rceil\right),
\end{equation}
where $o$ is the desired output size. Similarly, the depth increases with the number
of additional applications of $f$ required:
\begin{equation}
    \mathrm{depth}_{\mathrm{circuit}}
    = 2 + n_b(2 + 6n_{\text{rounds}})
      + 6\left\lceil\frac{o-r}{r}\right\rceil n_{\text{rounds}}.
\end{equation}

Since the multiplicative constants in these formulas are artefacts of the Reifier
compiler,
the asymptotic scaling laws for the standard Keccak construction are
\begin{align}
    \mathrm{width}_{\mathrm{circuit}}
        &= O\!\left(\max\!\left(11 \cdot 25w + p - r,\;
                               25w + r\left\lceil\tfrac{o-r}{r}\right\rceil\right)\right), \\[4pt]
    \mathrm{depth}_{\mathrm{circuit}}
        &= O\!\left(n_b \cdot n_{\text{rounds}}
                   + \left\lceil\tfrac{o-r}{r}\right\rceil\right).
\end{align}
These formulas were verified against the compiled circuits; Table~\ref{tab:keccak-arbitrary-formula-check} and Figures~\ref{fig:keccak_depth}, ~\ref{fig:keccak_width}
confirm that the predicted and actual depth and width values agree exactly across
all tested configurations.

\subsection{Merkle--Damg\aa rd Construction Experiments}
\label{subsec:md}

In these experiments we use $0^*$ padding to fill in the last block to have the same size as the previous blocks. Moreover,
we used a toy Keccak function $f$ with $w=1$, $n_{\text{rounds}}=1$, $r=12$ as the
compression function. The choice of these functions also sets the chain
length and the block size in these constructions to be the same:
$b_{\mathrm{size}} = c_{\mathrm{size}}$. 

\subsubsection{Layer-Level Inspection}
\label{subsubsec:md-circuit}

The first experiment was conducted on an input of $\mathit{in}=8$ bits which was
split into $N_b=2$ blocks of $N_{\mathrm{size}}=4$ bits. The compiled circuit can be found in
\texttt{md\_seq\_keccak\_N8\_M2\_logw1\_round1.txt}~\cite{github}.
Figure \ref{fig:Merkle-Damgard experimet} in Appendix \ref{app:MD_experiments} presents a simplified version of the circuit.

We see that per block we get depth $2+1+6n_{\text{rounds}}$: 2 from \textsc{XOR}-ing into the chain bits,
one from creating the Keccak state and 6 for the Keccak itself. This total
represents the depth of the compression function. Let $\mathrm{depth}_c$ denote the depth of compression function $c$ and in our case $c=f$ so
$\mathrm{depth}_c = 2+1+6 \cdot 1 = 9$. We can further generalise this to the depth
of the whole construction
\begin{equation}
    \mathrm{depth}_{MD_c} = 1 + \mathrm{depth}_c \cdot N_b + 1,
\end{equation}
where 1 comes from the initialisation of the chain value and the last 1 comes from
the output layer, where the input layer is excluded from the count.

Let $\mathrm{width}_c$ denote the width of a
compression function $c$, so in our case ($c=f$) we have $\mathrm{width}_c = 11 \cdot 25w$.
As noted above, all unabsorbed bits must be carried through the circuit at all times, which increases the width of the construction. We get the highest width at the
first pass of the compression function, when we only absorb one block and may have
multiple blocks to be absorbed, giving us
\begin{equation}
    \mathrm{width}_{MD_c} = \mathrm{width}_c + (N_b - 1) \cdot N_{\mathrm{size}}.
\end{equation}

Since the constants represent artefacts of the Reifier compiler we get the final
scaling laws for the MD construction to be
\begin{align}
    \mathrm{width}_{MD_c} &= O\!\left(\mathrm{width}_c + (N_b-1)\cdot N_{\mathrm{size}}\right), \\
    \mathrm{depth}_{MD_c} &= O\!\left(\mathrm{depth}_c \cdot N_b\right).
\end{align}
These formulas were verified against the compiled circuits; Table~\ref{tab:md-sequential-formula-check} and Figures~\ref{fig:md_depth_scaling}, ~\ref{fig:md_width_scaling}
confirm that predicted and actual depth and width values agree exactly across all
tested configurations.

\subsection{Merkle Tree Experiment}
\label{subsec:merkle-tree}

Due to
inconsistent generation of the circuits when padding was needed, experiments were restricted to inputs requiring no padding. Generalisation to padded inputs is left for further work. We used a toy Keccak function
$f_{\mathrm{leaves}}$ with $w=1$, $n_{\text{rounds}}=1$, $r=12$ as a compression function for
the leaves. Similarly, we define
$f_{\mathrm{nodes}} = f_{\mathrm{leaves}}(\mathrm{child}_{\mathrm{left}} \oplus
\mathrm{child}_{\mathrm{right}})$ to be the compression function for the nodes.
We used binary trees in our experiments. 

\subsubsection{Layer-Level Inspection}
\label{subsec:mt-circuit}

The first experiment was performed on an input of $\mathit{in}=8$ bits which was
split into $N_{\mathrm{in}}=2$ blocks of $b_{\mathrm{size}}=4$ bits. They were
later hashed using $f_{\mathrm{leaves}}$ and the resulting hashes were stored as the leaf values. These leaf values were used by their parent to compute
$f_{\mathrm{nodes}}(f_{\mathrm{leaves}}(\mathrm{block}_1) \oplus
f_{\mathrm{leaves}}(\mathrm{block}_2))$. The compiled circuit can be found in
\texttt{merkle\_tree\_keccak\_N8\_L2\_P2\_logw1\_round1.txt}~\cite{github}. Figure \ref{fig:MTexperiment} in Appendix \ref{app:MT_experiments} presents a simplified version of the circuit.
In total we have 17 layers, if we ignore the input layer: seven for $f_{\mathrm{leaves}}$, nine for
$f_{\mathrm{nodes}}$, and one for output. Generalising, for a toy Keccak function
$g$ with parameters $w$ and $n_{\text{rounds}}$, the depth formula is
\begin{equation}
    \mathrm{depth}_{\mathrm{circuit}}
    = \log_2(n_{\mathrm{leaves}}) \cdot (3 + 6n_{\text{rounds}}) + 2 + 6n_{\text{rounds}}.
\end{equation}

On the width of the circuit, we see that we perform the same operation in parallel
for multiple nodes. The widest layer is the first layer which computes the XOR of
the $\theta$ function inside $f$. We perform the most operations in parallel on
the leaves. Hence we get
\begin{equation}
    \mathrm{width}_{\mathrm{circuit}} = n_{\mathrm{leaves}} \cdot 11 \cdot 25w.
\end{equation}

In general, for some compression function $c$ with depth $\mathrm{depth}_c$ and
width $\mathrm{width}_c$ we have
\begin{align}
    \mathrm{width}_{\mathrm{circuit}} &= O(n_{\mathrm{leaves}} \cdot \mathrm{width}_c), \\
    \mathrm{depth}_{\mathrm{circuit}} &= O(\mathrm{depth}_c \cdot (1 + \log_2 n_{\mathrm{leaves}})),
\end{align}
as we perform the same function for each level in the tree. These formulas were
verified against the compiled circuits; Table~\ref{tab:mt-formula-check} and Figures~\ref{fig:mt_depth}, ~\ref{fig:mt_width} confirm
that predicted and actual depth and width values agree exactly across all tested
configurations.

\section{Mapping Circuits to Transformers}
\label{sec:mapping}

Throughout this section, we restrict our attention to the encoder stack of the transformer, consisting of alternating attention and FFN sublayers acting on a fixed-length token sequence.

\subsection{No attention mapping}
The simplest mapping ignores the attention mechanism entirely, reducing the transformer to its FFN sublayers. Since each FFN layer only mixes information within a single token, we store the entire input in a single token provided as input, where each token entry represents one wire value at a given circuit layer. Let $d$ be the dimension of the token. The most information needed at any time is the width of the circuit, so to store all the values we impose
\begin{equation}\
    d \;\geq\; \mathrm{width}_{\mathrm{circuit}},
\end{equation}
which also subsumes the condition $d \geq \text{input size}$ since the input layer is itself a circuit layer. Regarding token count, one token is enough to encode the input and we use the same token to store any intermediate result, giving $T = 1$. Furthermore, each FFN simulates the gate computations at one circuit layer, requiring $\mathrm{depth}_{\mathrm{circuit}}$ FFN sublayers, hence the transformer has depth $O\!\left(\mathrm{depth}_{\mathrm{circuit}}\right)$.

Each FFN layer directly corresponds to a SwiGLU layer in the MLP generated by the Reifier compiler. Hence, the maximum FFN width equals the maximum width of the generated MLP.

We therefore reduce the transformer to the MLP generated by the Reifier library, adding no additional structure or expressiveness. This mapping establishes a lower bound for the depth of any transformer mapping
\begin{equation}
    \mathrm{depth}_T \;\geq\; \mathrm{depth}_{\mathrm{circuit}}
\end{equation}
regardless of how attention and tokens are arranged, since no single sublayer, attention or FFN, can evaluate more than one circuit layer simultaneously.

\subsection{Tokens-as-gates}

The mappings in this section follow Merrill et al.'s~\cite{merrill2022saturated} count-based threshold gate
definition $\theta^{\geq k}$, which is equivalent in expressivity to the signed-weight
formulation of Hajnal et al.~\cite{hajnal1993threshold} used in Section~\ref{sec:background}. We define
$T_{\text{gate}}$ as the transformer in which each token represents a single gate in
the given circuit. The number of input bits is far smaller than the number
of total gates in a circuit, so we define scratch tokens to be additional tokens
initialised to store the values of intermediate and output gates. Hence we
have $T = T_{\text{in}} + T_{\text{scratch}} = G$, where $G$ is the total number of
gates within the circuit. 

We then propose a mapping to a single-head attention transformer. Attention simulates the circuit wiring, while the FFN simulates the gate computation. Without loss
of generality, let token $t_i$ represent gate $g_i$ and $x_i \in \mathbb{R}^{T+2}$ be
the embedded vector of $t_i$. We use the first $T$ entries of $x_i$ as a one-hot
encoding of $g_i$. Entry $x_{T+1}$ represents the current value of the gate, which is
the value of the input bit for input gates and $0$ at initialisation for all other
gates. Constant gates, which Reifier spawns as nodes with no predecessors and a fixed
value, require no special treatment. Their value is hardcoded in the token embedding from
the start. Entry $x_{T+2}$ represents the threshold of the gate. We depart from Merrill et al.'s~\cite{merrill2022saturated} saturated attention by \emph{assuming} unnormalised hard attention: we set $a_{i,j} = 1$ for each predecessor. This is an idealising assumption, since a standard softmax sublayer cannot produce these exact $\{0,1\}$ weights. We choose the attention matrices such that all predecessor dot products score exactly 1 and all
non-predecessors score exactly 0, so the
attention output is the sum of predecessor values. The FFN then compares this sum
directly against the integer threshold $\theta$ stored in $x_i[T+2]$, avoiding the
need for a normalised threshold.

The attention output is zero in all entries except the $(T+1)$th, which holds the sum of predecessor gate values. We retrieve the one-hot
encoding and the threshold using the skip connection before the attention layer. This
gives a vector with the same entries as the input to the layer, but with
the sum of predecessor values in the $(T+1)$-th entry. Only the gates at the current circuit layer are relevant. We use the same
attention matrices for all attention sublayers. These overwrite previously computed gate values, but this
does not affect correctness because Reifier produces a strictly layered circuit in which each gate's
predecessors lie in the immediately preceding layer, hence the $t$-th block computes the correct value of every
gate at circuit layer $t$, regardless of the possibly stale values held by gates at other layers. The output
is then read from the tokens corresponding to output gates.

To ensure the attention output equals the sum of predecessor values, we require $a_{i,j} \leq 1$, which is guaranteed by the following matrix definitions. Let $Q \in \mathbb{R}^{T \times (T+2)}$,
$K \in \mathbb{R}^{T \times (T+2)}$, $V \in \mathbb{R}^{(T+2) \times (T+2)}$ and
define them as follows:
\begin{equation} \label{eq: Q}
    Q_{i,j} = \begin{cases} 1 & \text{if } i = j,\; i \leq T \\ 0 & \text{otherwise} \end{cases}
\end{equation}
\begin{equation} \label{eq: K}
    K_{i,j} = \begin{cases} 1 & j \in \mathrm{pred}(i),\; j \leq T \\ 0 & \text{otherwise} \end{cases}
\end{equation}
\begin{equation} \label{eq: V}
    V_{i,j} = \begin{cases} 1 & i = j = T+1 \\ 0 & \text{otherwise} \end{cases}
\end{equation}

We present the analysis of this construction in Appendix \ref{app:tokens}.

Each encoder sublayer
simulates one layer of the circuit, giving
\begin{equation}
    \mathrm{depth}_{\text{encoder}} \;=\; O\!\left(\mathrm{depth}_{\text{circuit}}\right)
\end{equation}
consistent with the lower bound established in the previous mapping.


\section{Discussion}
\label{sec:discussion}

The experiments confirm that all three constructions admit threshold-circuit representations, whose scaling laws directly constrain the transformer architectures required to implement them.
The corresponding transformers have encoder stacks whose depth is
proportional to the depth of the circuit, while the FFN width does not depend on the parameters of the cryptographic constructions or number of input blocks
for the tokens-as-gates mapping. In the no-attention case the FFN width does depend on the cryptographic construction and the input size, since each FFN corresponds to one SwiGLU layer in the compiled MLP.
 
The clearest comparison can be made between the MDC and MT, since they rely on similar compression functions and
mechanisms. 
MD produces deeper but narrower circuits, while MT produces shallower but wider ones.
A depth--width trade-off is visible across these two constructions,
while the overall computational work remains of similar order, both requiring $O(N_b \cdot n_c)$ gates, where $n_c$ denotes the number of gates required to compute the compression functions and $N_b$ the number of input blocks. 
 
We also observe a trade-off between the two proposed transformer mappings.
The no-attention mapping does not exploit the expressive power of
the attention mechanism, resulting in disproportionately wide FFN layers whose width grows
linearly with the circuit width.
By contrast, the tokens-as-gates mapping has constant width FFNs, but requires more tokens and introduces attention routing absent from the no-attention mapping. 


\section{Conclusion and Further Work}
\label{sec:conclusion}

Building on the $TC^0$ characterisation of saturated transformers~\cite{merrill2022saturated} and the cryptographic randomness result of~\cite{motwani2025secretcollusion}, we mapped three families of hash functions to threshold circuits and proposed concrete transformer implementations, deriving depth and FFN width bounds for each.
 
All three constructions obey verified scaling laws, confirmed against compiled circuits across all tested configurations.
Two transformer mappings were proposed:
no-attention baseline and tokens-as-gates mapping.
Both satisfy the depth lower bound $O(\mathrm{depth}_{\mathrm{circuit}})$,
while differing in their token and embedding-dimension requirements.
 
Several important caveats apply.
The mappings are constructive and have not been verified empirically by training,
whether gradient descent can learn to implement these constructions is an open
question.
The tokens-as-gates mapping assumes unnormalised hard attention, introducing a gap between the theoretical model
and transformers used in practice.
Furthermore, the experiments were conducted on toy instances of the
constructions, and the depth and width formulas contain constants that are
artefacts of the Reifier compiler~\cite{reifier}; a different compiler
could produce tighter bounds.
Whether the scaling laws hold without qualification for the full Keccak
permutation as used in SHA-3, or for MD constructions at production scale,
remains to be confirmed.
 
Notwithstanding these limitations, this work provides the first systematic
circuit-level analysis of Keccak functions, MD constructions,
and MT in the context of transformer expressivity, and proposes concrete
mappings that serve as a firm basis for further mathematical and empirical
investigation.
 
Several directions merit further investigation: constructing explicit transformers with specified attention heads and embedding dimensions, testing whether gradient descent can learn these functions, and developing alternative mappings such as tokens-as-states or tokens-as-values.

We hope to develop future frontier model benchmark evaluations grounded in circuit-theoretic insights into transformer computational capacity.


\bibliography{references}

\appendix

\section{Keccak Operation Definitions}
\label{app:keccak_def}

The state $a$ defined in Section \ref{bg:keccak} has length $25w$ where
$w \in \{1, 2, 4, 8, 16, 32, 64\}$, depending on the version of Keccak that is used. Capacity has length $c$ and rate has length $r$.
 
Define $GF(p)$ for $p$ a prime number as the Galois Field with $p$ elements, where addition and multiplication are performed modulo $p$. Moreover, we define $GF(p)[X]$ as the polynomials with
coefficients over $GF(p)$.

The padding function is defined as $pad = 10^*1$ and it adds at least 2 bits to the message
and at most $r+1$. Note that the state $a$ is a
3 dimensional array $a[5][5][w]$ with binary content $a[x][y][z] \in GF(2)$, where
$w = 2^l, l \in \{0, 1, 2, 3, 4, 5, 6\}$ is another parameter for the sponge construction. We say that each entry $a[x][y]$ represents a word and $w$ is the word size. All the operations will take place in $GF(2)$ unless stated otherwise.
We write $s[w(5y+x)+z] = a[x][y][z]$, $x, y \in \mathbb{Z}_5$, $z \in \mathbb{Z}_w$, the
state as a one dimensional array. For Keccak, the 5 permutations that make $f$ are performed
$n_{\text{rounds}}$ times. So we can write
$f = (\iota \circ \chi \circ \pi \circ \rho \circ \theta)^{[n_{\text{rounds}}]}$. Now we can
define these 5 operations.
 
$\theta$ computes
\begin{equation}
    a[x][y][z] \leftarrow a[x][y][z] + \sum_{y'=0}^{4} a[x-1][y'][z] + \sum_{y'=0}^{4} a[x+1][y'][z-1]
\end{equation}
 
$\rho$ computes
\begin{equation}
    a[x][y][z] \leftarrow a[x][y][z-(t+1)(t+2)/2]
\end{equation}
where $0 \leq t < 24$ and
$\begin{pmatrix} 0 & 1 \\ 2 & 3 \end{pmatrix}^t \begin{pmatrix} 1 \\ 0 \end{pmatrix} =
\begin{pmatrix} x \\ y \end{pmatrix} \in GF(5)^{2 \times 2}$,
or $t = -1$ for $x = y = 0$. $\rho$ bitwise rotates each word with a different triangular number.
 
$\pi$ computes $a[x][y] \leftarrow a[x'][y']$, with
\begin{equation}
    \begin{pmatrix} x \\ y \end{pmatrix} =
    \begin{pmatrix} 0 & 1 \\ 2 & 3 \end{pmatrix}
    \begin{pmatrix} x' \\ y' \end{pmatrix}
\end{equation}
$\pi$ permutes the 25 words in a fixed pattern.
 
$\chi$ computes
\begin{equation}
    a[x] \leftarrow a[x] + (a[x+1]+1)\,a[x+2]
\end{equation}
This operation bitwise combines along the rows. This is the only non-linear operation in
Keccak.
 
$\iota$ computes $a \leftarrow a + RC[i_r]$. $RC[i_r]$ is a round constant computed using
$RC[i_r][0][0][2^j-1] = rc[j+7i_r]$, $\forall\, 0 \leq j \leq l$. All the other terms
$RC[i_r][x][y][z] = 0$. Moreover, they define
$rc[t] = (x^t \bmod x^8+x^6+x^5+x^4+1) \bmod x$ in $GF(2)[X]$. This operation breaks the
symmetry preserved by the other operations. Since $RC[i_r][x][y][z]=0, (x,y) \neq (0, 0)$, only lane $a[0][0]$ is affected.
 
The number of rounds is $n_{\text{rounds}} = 12 + 2l$.

Now we present Algorithms 2 and 3 mentioned in Section \ref{bg:keccak}.

\begin{algorithm}[h]
\caption{Absorb phase}
\begin{algorithmic}[1]
\Procedure{Absorb}{$x[1 \ldots n]$, $f$}
    \For{$i \gets 1$ \textbf{to} $n$}
        \State $a[0 \ldots r-1] \gets a[0 \ldots r-1] \oplus x[i]$
        \State $a \gets f(a)$
    \EndFor
\EndProcedure
\end{algorithmic}
\end{algorithm}
 
\begin{algorithm}[h]
\caption{Squeeze phase}
\begin{algorithmic}[1]
\Procedure{Squeeze}{$\text{out\_size}$, $f$}
    \State $\text{result} \gets \varepsilon$
    \While{$|\text{result}| < \text{out\_size}$}
        \If{$|\text{result}| + r \leq \text{out\_size}$}
            \State $\text{result} \gets \text{result} \;\|\; a[0 \ldots r-1]$
        \Else
            \State $\text{result} \gets \text{result} \;\|\; a[0 \ldots \text{out\_size} - |\text{result}| - 1]$
        \EndIf
        \State $a \gets f(a)$
    \EndWhile
    \State \Return $\text{result}$
\EndProcedure
\end{algorithmic}
\end{algorithm}
\section{Detailed Keccak Experiments}
\label{app:keccak_experiments}

\begin{figure}[h]
    \centering
    \includegraphics[width=0.5\linewidth]{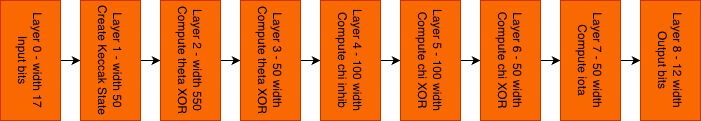}
    \caption{Circuit compiled for Keccak function with $w=1$ and $n_{\text{rounds}}=1$}
    \label{fig:keccak1}
\end{figure}

In the experiment presented in Section \ref{subsubsec:keccak-circuit} Layer 1 prepares the internal state of the sponge construction,
initialising the remaining state bits to zero using constant nodes. Layer 2 and 3 both implement the $\theta$ function for which we need to compute a \textsc{XOR}.

In the $\theta$ function we compute a XOR over 11 elements,
so the layer which computes the first part of the XOR ends up having $11 \cdot 25w$
nodes: one set of $t=11$ threshold gates per element of the $25w$-element state.
The subsequent layer performs only one threshold per 11 previous nodes, so
it has width $25w$.

Functions $\rho$ and $\pi$ only compute permutations and they are simulated through the wiring of the circuit, contributing with no layers to the compiled result.

Layer 3 computes what the Reifier library calls the \emph{inhib} part of the
operation, namely $\lnot a[x+1] \;\&\; a[x+2]$, using the threshold gate
$\mathbf{1}[-a[x+1] + a[x+2] \geq 1]$. This part requires one threshold per element
of the state ($25w$ nodes), but also introduces copies of the values from the previous
layer to enable the XOR computation in the following two layers ($25w$
nodes), giving a total width of $2 \cdot 25w$.

Layer 7 requires only one layer, because it is a
two-element \textsc{XOR} with the constant $RC[i_r]$ which is hardwired into the gate's threshold function.
Notably, experiments show that this layer is omitted entirely when $RC[i_r] = 0$,
since the \textsc{XOR} is then redundant. This is an artefact of the Reifier compiler.

The final layer is the output layer, which outputs the $r$ bits of the Keccak state.

\begin{table}[h]
\centering
\caption{Deviation from the linear prediction $\mathrm{depth}_{\mathrm{pred}}(n_{\text{rounds}})=6n_{\text{rounds}}+2$. Positive $\Delta$ means more layers than predicted. Deviations take place because of $RC[i_r]=0$}
\label{tab:keccak-pred-vs-obs}
\begin{tabular}{r r r r l}
\toprule
$\log w$ & $n_{\text{rounds}}$ & $\mathrm{depth}_{\mathrm{obs}}$ & $\Delta = \mathrm{depth}_{\mathrm{obs}}-\mathrm{depth}_{\mathrm{pred}}$ & notes \\
\midrule
2 & 13 & 79  & -1 & reported: lost 1 \\
2 & 19 & 114 & -2 & reported: lost 2 \\
3 & 15 & 91  & -1 & reported: lost 1 \\
4 & 24 & 146 & 0  & matches $\mathrm{depth}_{\mathrm{pred}}(24)=146$ \\
\bottomrule
\end{tabular}
\end{table}

\begin{table}[h]
\centering
\caption{Formula verification for arbitrary Keccak input and output sizes.}
\label{tab:keccak-arbitrary-formula-check}
\begin{tabular}{cccccccc}
\toprule
$\log w$ & Rate & Blocks & Rounds & Exp.\ depth & Actual depth & Exp.\ width & Actual width \\
\midrule
1 & 25 & 1 & 1 & 10 & 10 & 550  & 550  \\
1 & 25 & 2 & 1 & 18 & 18 & 575  & 575  \\
1 & 25 & 4 & 1 & 34 & 34 & 625  & 625  \\
1 & 25 & 1 & 2 & 16 & 16 & 550  & 550  \\
1 & 25 & 2 & 2 & 30 & 30 & 575  & 575  \\
1 & 25 & 4 & 2 & 58 & 58 & 625  & 625  \\
2 & 50 & 1 & 1 & 10 & 10 & 1100 & 1100 \\
2 & 50 & 2 & 1 & 18 & 18 & 1150 & 1150 \\
2 & 50 & 4 & 1 & 34 & 34 & 1250 & 1250 \\
2 & 50 & 1 & 2 & 16 & 16 & 1100 & 1100 \\
2 & 50 & 2 & 2 & 30 & 30 & 1150 & 1150 \\
2 & 50 & 4 & 2 & 58 & 58 & 1250 & 1250 \\
\bottomrule
\end{tabular}
\end{table}

\begin{figure}[h]
    \centering
    \includegraphics[width=1\linewidth]{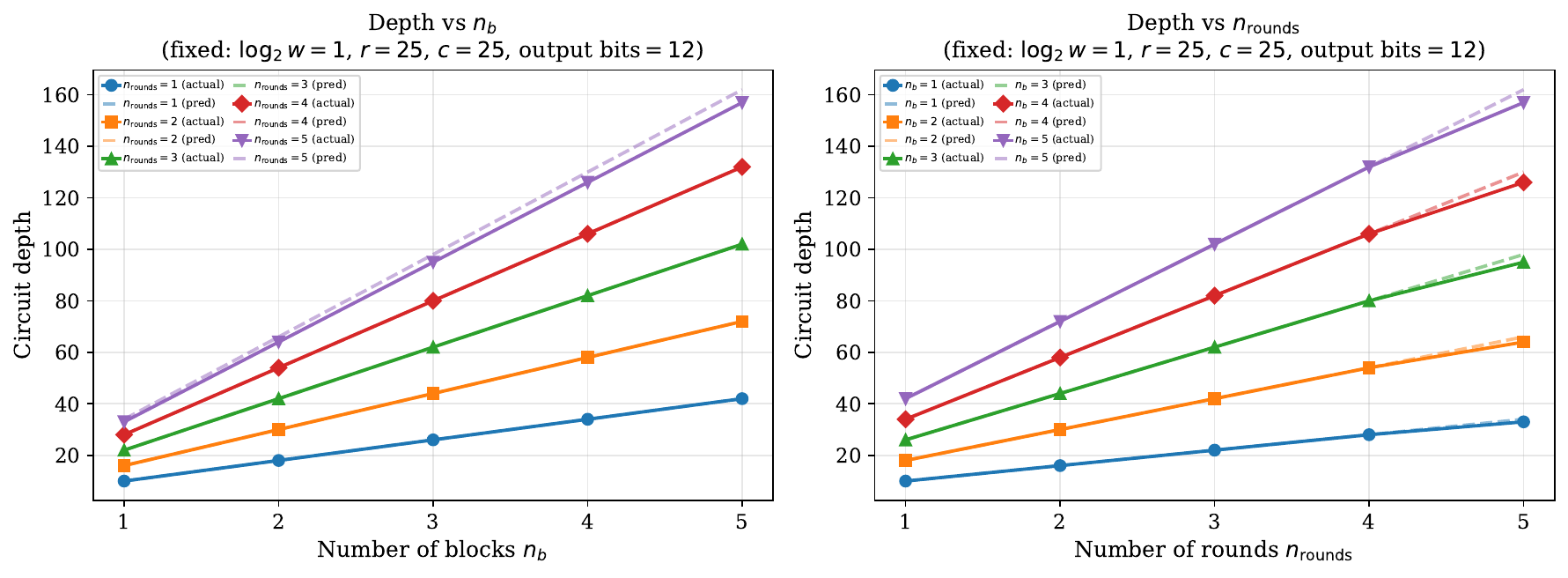}
    \caption{Plot for Keccak depth scaling. Differences between predicted and actual values come from round constant which are 0}
    \label{fig:keccak_depth}
\end{figure}

\begin{figure}[h]
    \centering
    \includegraphics[width=1\linewidth]{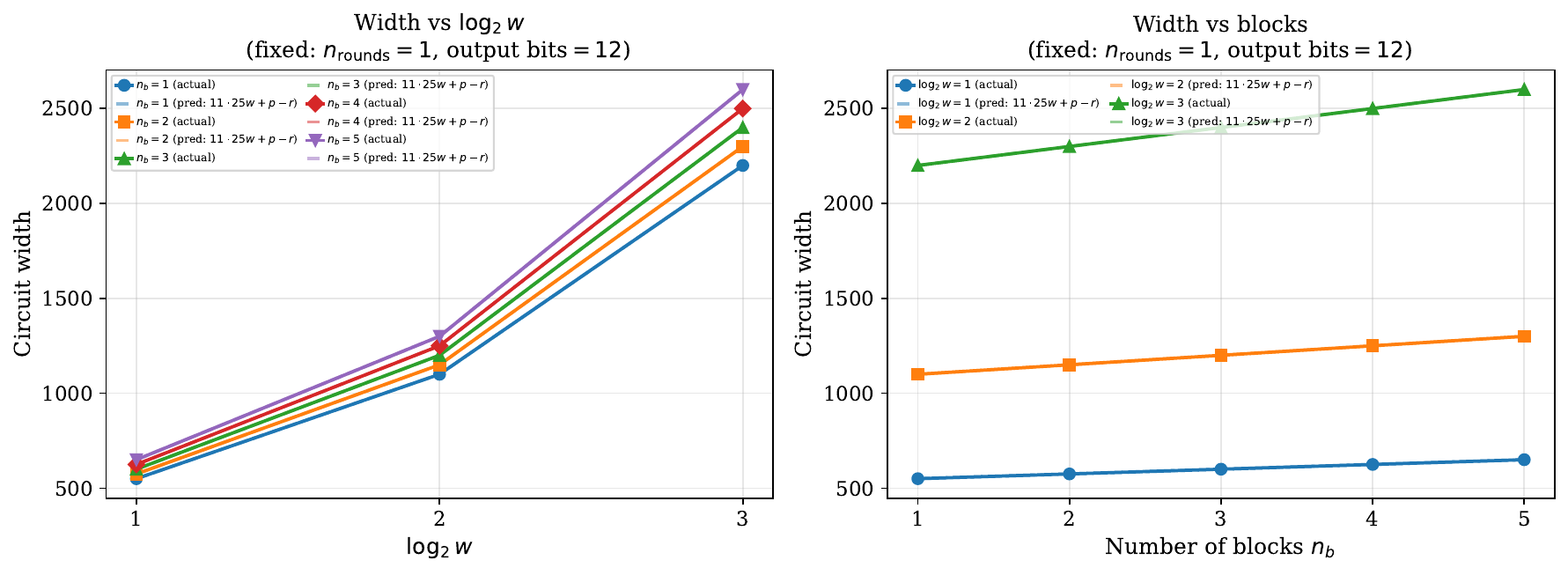}
    \caption{Plot for Keccak width scaling}
    \label{fig:keccak_width}
\end{figure}

\section{Detailed MD Experiments}
\label{app:MD_experiments}

\begin{figure}[h]
    \centering
    \includegraphics[width=0.65\linewidth]{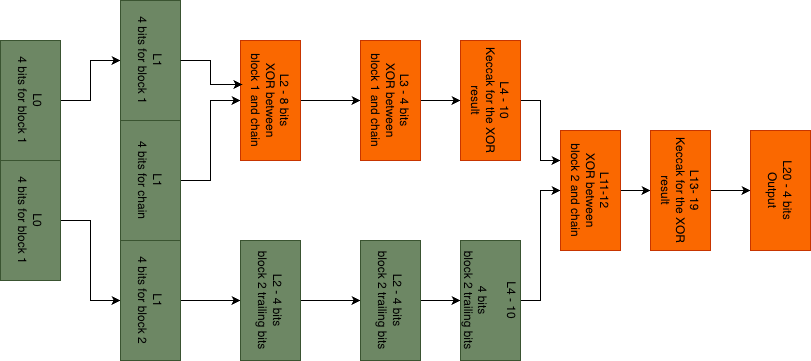}
    \caption{Circuit compiled for MD with compression function Keccak with $w=1$ and $n_{\text{rounds}}=1$}
    \label{fig:Merkle-Damgard experimet}
\end{figure}

\begin{figure}[h]
    \centering
    \includegraphics[width=0.5\linewidth]{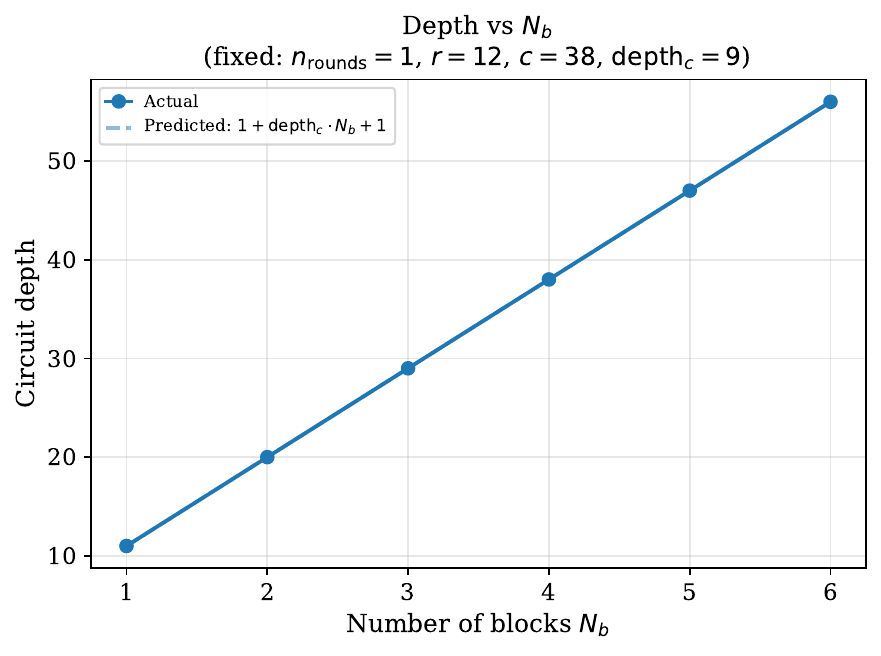}
    \caption{Plot for MD depth scaling}
    \label{fig:md_depth_scaling}
\end{figure}

\begin{figure}[h]
    \centering
    \includegraphics[width=1\linewidth]{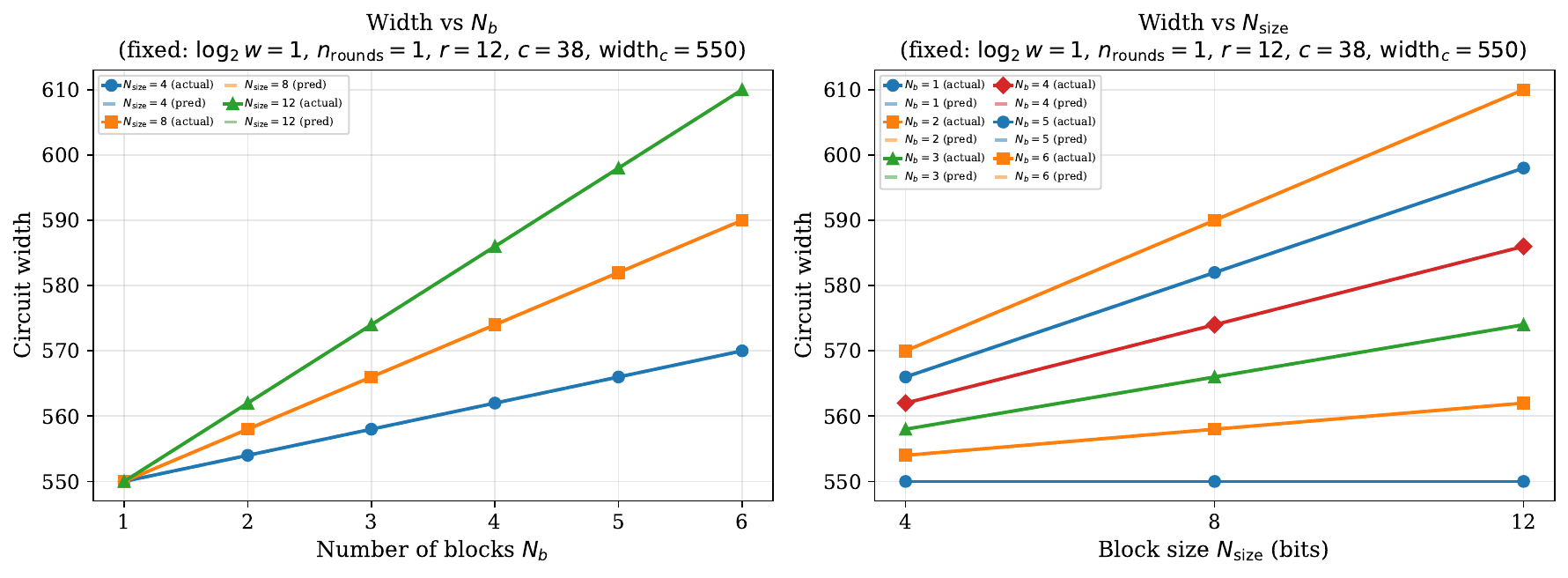}
    \caption{Plot for MD width scaling}
    \label{fig:md_width_scaling}
\end{figure}

Below we have the circuit (Fig. \ref{fig:Merkle-Damgard experimet}), tables (Table \ref{tab:md-sequential-formula-check}) and graphs (Graphs \ref{fig:md_depth_scaling}, \ref{fig:md_width_scaling}) of our experiment results. Notably, the width increases by four when performing the Keccak function, as we need to keep the bits of the input that will subsequently be absorbed by the construction. The XOR computed after the first Keccak instance is part of the absorption of the next block.

\begin{table}[h]
\centering
\caption{Formula verification for the sequential Merkle--Damg{\aa}rd construction.}
\label{tab:md-sequential-formula-check}
\small
\resizebox{\textwidth}{!}{%
\begin{tabular}{cccccccccc}
\toprule
$\log w$ & Rate & Block bits & Chain bits & Blocks & Rounds
& Exp.\ depth & Actual depth & Exp.\ width & Actual width \\
\midrule
1 & 12 & 4 & 4 & 1 & 1 & 11 & 11 & 550  & 550  \\
1 & 12 & 4 & 4 & 2 & 1 & 20 & 20 & 554  & 554  \\
1 & 12 & 4 & 4 & 3 & 1 & 29 & 29 & 558  & 558  \\
1 & 12 & 4 & 4 & 4 & 1 & 38 & 38 & 562  & 562  \\
1 & 16 & 8 & 8 & 1 & 1 & 11 & 11 & 550  & 550  \\
1 & 16 & 8 & 8 & 2 & 1 & 20 & 20 & 558  & 558  \\
1 & 16 & 8 & 8 & 3 & 1 & 29 & 29 & 566  & 566  \\
1 & 16 & 8 & 8 & 4 & 1 & 38 & 38 & 574  & 574  \\
1 & 12 & 4 & 4 & 1 & 2 & 17 & 17 & 550  & 550  \\
1 & 12 & 4 & 4 & 2 & 2 & 32 & 32 & 554  & 554  \\
1 & 12 & 4 & 4 & 3 & 2 & 47 & 47 & 558  & 558  \\
1 & 12 & 4 & 4 & 4 & 2 & 62 & 62 & 562  & 562  \\
1 & 16 & 8 & 8 & 1 & 2 & 17 & 17 & 550  & 550  \\
1 & 16 & 8 & 8 & 2 & 2 & 32 & 32 & 558  & 558  \\
1 & 16 & 8 & 8 & 3 & 2 & 47 & 47 & 566  & 566  \\
1 & 16 & 8 & 8 & 4 & 2 & 62 & 62 & 574  & 574  \\
2 & 12 & 4 & 4 & 1 & 1 & 11 & 11 & 1100 & 1100 \\
2 & 12 & 4 & 4 & 2 & 1 & 20 & 20 & 1104 & 1104 \\
2 & 12 & 4 & 4 & 3 & 1 & 29 & 29 & 1108 & 1108 \\
2 & 12 & 4 & 4 & 4 & 1 & 38 & 38 & 1112 & 1112 \\
2 & 16 & 8 & 8 & 1 & 1 & 11 & 11 & 1100 & 1100 \\
2 & 16 & 8 & 8 & 2 & 1 & 20 & 20 & 1108 & 1108 \\
2 & 16 & 8 & 8 & 3 & 1 & 29 & 29 & 1116 & 1116 \\
2 & 16 & 8 & 8 & 4 & 1 & 38 & 38 & 1124 & 1124 \\
2 & 12 & 4 & 4 & 1 & 2 & 17 & 17 & 1100 & 1100 \\
2 & 12 & 4 & 4 & 2 & 2 & 32 & 32 & 1104 & 1104 \\
2 & 12 & 4 & 4 & 3 & 2 & 47 & 47 & 1108 & 1108 \\
2 & 12 & 4 & 4 & 4 & 2 & 62 & 62 & 1112 & 1112 \\
2 & 16 & 8 & 8 & 1 & 2 & 17 & 17 & 1100 & 1100 \\
2 & 16 & 8 & 8 & 2 & 2 & 32 & 32 & 1108 & 1108 \\
2 & 16 & 8 & 8 & 3 & 2 & 47 & 47 & 1116 & 1116 \\
2 & 16 & 8 & 8 & 4 & 2 & 62 & 62 & 1124 & 1124 \\
\bottomrule
\end{tabular}}
\end{table}

\section{Detailed MT Experiments}
\label{app:MT_experiments}

\begin{figure}[h]
    \centering
    \includegraphics[width=0.5\linewidth]{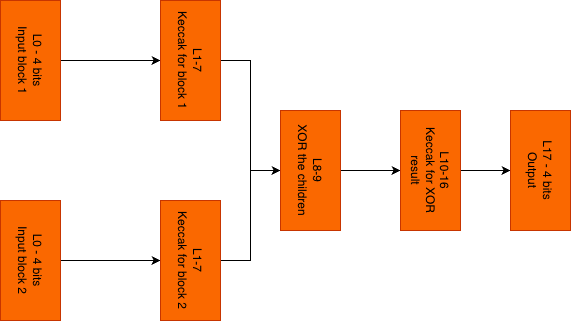}
    \caption{Circuit compiled for MT with compression function Keccak with $w=1$ and $n_{\text{rounds}}=1$}
    \label{fig:MTexperiment}
\end{figure}

In the circuit presented in Figure \ref{fig:MTexperiment}
we observe that the first layer represents the input and has $\mathrm{width}=8$.
In the second layer we start computing $f_{\mathrm{leaves}}$ for each leaf in
parallel by creating the Keccak state. Since each Keccak state has 50 entries,
two leaves being processed in parallel gives $2 \cdot 50 = 100$ nodes. The next
six layers perform the Keccak toy function $f$. Each of these layers has a width
twice as large as its corresponding layer in the circuit representation of $f$.
Layers 9 and 10, the first layers after $f_{\mathrm{leaves}}$ has been computed,
XOR the previous results and thus start computing $f_{\mathrm{nodes}}$. The
following seven layers consist of creating the Keccak state and then computing
$f_{\mathrm{nodes}}$. These layers have the same width as the corresponding layer
in the circuit representation of $f$. The last layer represents the output. 

\begin{table}[h]
\centering
\caption{Formula verification for the Merkle tree construction.}
\label{tab:mt-formula-check}
\begin{tabular}{cccccccc}
\toprule
$\log w$ & $w$ & Rounds & Leaves
& Exp.\ depth & Actual depth & Exp.\ width & Actual width \\
\midrule
1 & 2 & 1 & 1 &  8 &  8 &  550 &  550 \\
1 & 2 & 1 & 2 & 17 & 17 & 1100 & 1100 \\
1 & 2 & 1 & 4 & 26 & 26 & 2200 & 2200 \\
1 & 2 & 1 & 8 & 35 & 35 & 4400 & 4400 \\
1 & 2 & 2 & 1 & 14 & 14 &  550 &  550 \\
1 & 2 & 2 & 2 & 29 & 29 & 1100 & 1100 \\
1 & 2 & 2 & 4 & 44 & 44 & 2200 & 2200 \\
1 & 2 & 2 & 8 & 59 & 59 & 4400 & 4400 \\
2 & 4 & 1 & 1 &  8 &  8 & 1100 & 1100 \\
2 & 4 & 1 & 2 & 17 & 17 & 2200 & 2200 \\
2 & 4 & 1 & 4 & 26 & 26 & 4400 & 4400 \\
2 & 4 & 1 & 8 & 35 & 35 & 8800 & 8800 \\
2 & 4 & 2 & 1 & 14 & 14 & 1100 & 1100 \\
2 & 4 & 2 & 2 & 29 & 29 & 2200 & 2200 \\
2 & 4 & 2 & 4 & 44 & 44 & 4400 & 4400 \\
2 & 4 & 2 & 8 & 59 & 59 & 8800 & 8800 \\
\bottomrule
\end{tabular}
\end{table}

\begin{figure}[h]
    \centering
    \includegraphics[width=0.5\linewidth]{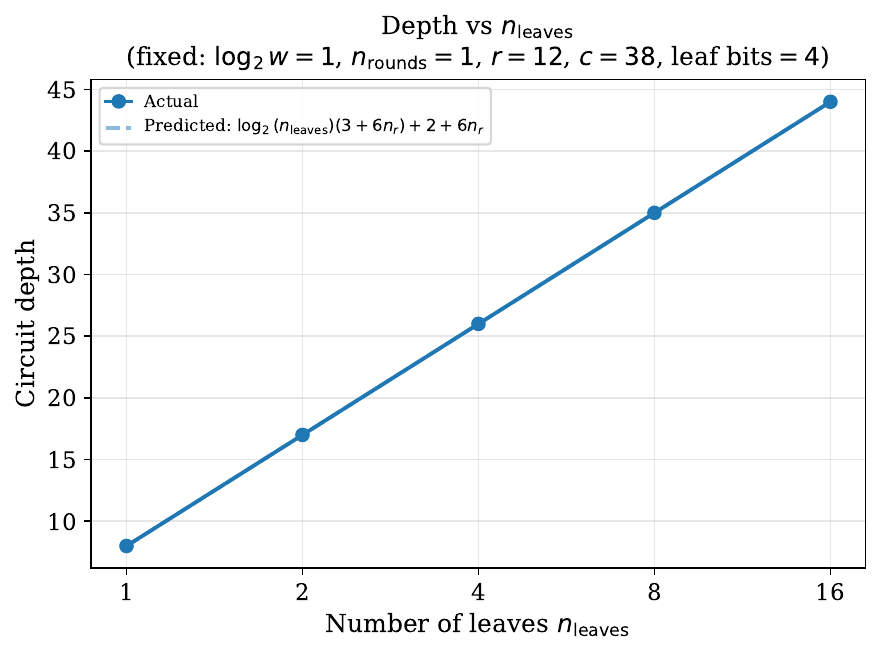}
    \caption{Plot for MT depth scaling}
    \label{fig:mt_depth}
\end{figure}

\begin{figure}[h]
    \centering
    \includegraphics[width=0.5\linewidth]{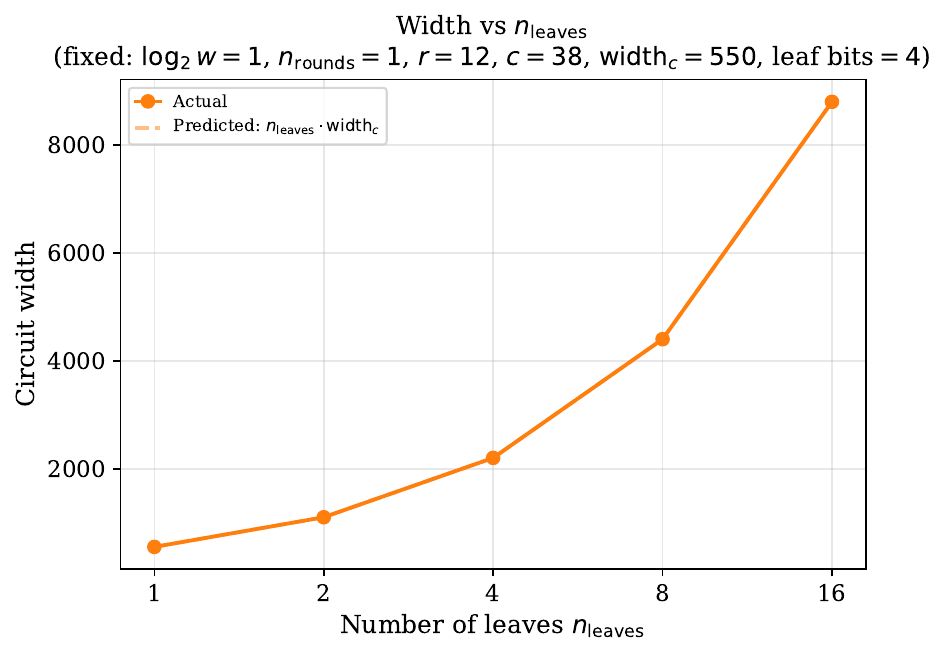}
    \caption{Plot for MT width scaling}
    \label{fig:mt_width}
\end{figure}

\section{Tokens-as-Gates Correctness}\label{app:tokens}

For the Tokens-as-Gates mapping we get the following flow of computation for the
attention sublayer:
\begin{align*}
    q_i &= Qx_i \\
    k_i &= Kx_i \\
    v_i &= Vx_i \\
    s_{i,j} &= q_i^T k_j \\
    a_{i,j} &= \begin{cases} 1 & \text{if } s_{i,j} = 1 \\ 0 & \text{otherwise} \end{cases} \\
    \mathrm{out}_i &= \sum_j a_{i,j}\, v_j
\end{align*}

Looking at equations \ref{eq: K}, \ref{eq: Q}, \ref{eq: V} note that $K$ is a fixed matrix. When applied to $x_j = e_j$, it returns a vector
whose $i$-th entry is 1 if gate $j$ is a predecessor of gate $i$, encoding gate $j$'s
successor structure. These choices ensure that after the attention sublayer we retrieve the sum of the values of the predecessor gates. The FFN then evaluates the threshold function using the gate value (entry $T+1$) and threshold (entry $T+2$).

An algebraic analysis for this would be, that for an input token $t_i$ which represents $g_i$ and is further represented by $x_i$ as an embedded vector, we have the following values for $q_i, k_i \text{ and } v_i$.

\begin{align*}
q_i &= Q x_i = \text{one-hot encoding of } g_i \\
k_j &= K x_j \\
k_{j,l} &
        = K_{l,j} + K_{l,T+1}\cdot(\text{gate value}) + K_{l,T+2}\cdot\theta
        = K_{l,j} \\
s_{i,j} &= q_i^{\top} k_j =
        \begin{cases}
          1 & \text{if } k_{j,i} = K_{i, j} = 1 \\
          0 & \text{otherwise}
        \end{cases} \\
v_{i,j} &= V_j \cdot x_i =
        \begin{cases}
          \text{gate value} & \text{if } j = T+1 \\
          0 & \text{otherwise}
        \end{cases}
\end{align*}

where $k_{i, j}$ is the $j^{\text{th}}$ entry of $k_i$ and $x_{i, l}$ is the $l^{\text{th}}$ entry of $x_i$.

We want to accumulate the sum of the values of the previous gates and we can do so if we have $a_{i, j} = 1$ for $j \in \text{pred}(i)$. Looking back at the computational flow, note that the output will be a vector with all entries zero except entry $T+1$ where we will have the sum of the values of the predecessors. We retrieve the one hot encoding and the threshold value by using the residual connections in the encoder block.

In our experiments, the NOT operation appears only within the $\chi$
function of the
Keccak construction. The circuit uses gates that compute
$(\lnot a[x+1] \mathbin{\&} a[x+2])$ as $-a[x+1] + a[x+2] \geq 1$. Thus, simple NOT gates are never required, but if needed, we could add a new dimension to the
token vector to signal to the FFN that a NOT gate is being computed, inverting the relevant bit.

\end{document}